\documentstyle[preprint,aps]{revtex}

\begin{document}

\begin{titlepage}

\hfill {\tt SOGANG-HEP 282/01}

\begin{center}
{\Large{{\bf Noncommutative field theory description of quantum
Hall skyrmions }}} \vskip 0.5cm \end{center}

\begin{center}
{Bum-Hoon Lee$^{1}$, Kyungsun Moon$^{2}$ and Chaiho Rim$^{3}$}\par
\end{center}
\begin{center}
{$^{1}$Department of Physics, Sogang University, C.P.O. Box 1142,
Seoul 100-611, Korea} \par {$^{2}$Department of Physics and IPAP,
Yonsei University, Seoul 120-749, Korea}\par {$^{3}$Department of
Physics, Chonbuk National University, Chonju 561-756, Korea}
\end{center}
\vskip 0.3cm
\begin{center}
{\bf ABSTRACT}
\end{center}
\begin{quotation}
We revisit the quantum Hall system with no Zeeman splitting energy
using the noncommutative field theory. We analyze the BPS
condition for the delta-function interaction near the filling
factor $\nu=1$. Multi-skyrmions are shown to saturate the BPS
bounds. The dimension of the moduli space of $k$ skyrmions is
$4k+2$. Advantage of the noncommutative field description is
demonstrated through the derivation of the effective nonlinear
$\sigma$ model Lagrangian.

---------------------------------------------------------------------\\
\noindent
$^1$bhl@ccs.sogang.ac.kr, $^2$kmoon@phya.yonsei.ac.kr\\
$^3$rim@phy.chonbuk.ac.kr\\
\noindent
\end{quotation}
\end{titlepage}

\newpage

\section{Introduction}

In recent years there have been much interests in the
noncommutative field theories motivated by the low energy
effective theory of the string theory in the presence of the
string background field $B_{\mu\nu}$ \cite{seibergwitten,jabbari}.
The noncommutative field theories have shown many interesting
properties. Especially, non-commutative solitons have been
suggested, which have quite different properties from those in the
ordinary field theories\cite{soliton,gms}.

It has also been known in condensed matter physics that the
electron system in the lowest Landau level in the strong magnetic
field such as the quantum Hall system can be understood on the
noncommutative space\cite{read}. Naturally many works deal with
the quantum Hall systems in terms of noncommutative field
theories\cite{bigatti2,gubser2,susskind01,myung} in a general
setting. Moreover, it was presented in \cite{pasquier,pasquier2}
that the skyrmionic excitation on a sphere is the soliton in the
noncommutative field theory. The presentation of the skyrmion is
very suggestive but seems not manifest in terms of the
noncommutative field theory, not to mention the dipole
description.

In this paper, we revisit the multi-skyrmions on the planar
quantum Hall system near the filling factor $\nu=1$, in the limit
of vanishing Zeeman splitting energy.  The energy of the system
is bounded below. The way how the multi-skyrmion
configuration in the lowest Landau level (LLL) saturates the
energy is more intriguing than that in the full two-dimensional
system. For the delta-function interaction, multi-skyrmion
solutions saturating this BPS bounds are explicitly constructed.

The quantum Hall system is also described in terms of the second
quantized field theory where the electron states are projected to the
LLL. Based on the Hartree-Fock approximation, the system is
constructed so that the potential is written in the manifest
noncommutative form.

In section 2, we review the single electron in the LLL using the
coherent state representation, emphasizing the noncommutativity
coming from the projection to the LLL. We also introduce the
$*$-product of functions that is convenient to deal with the
noncommutative properties of the LLL system. In section 3, in
terms of the second quantized fields on the LLL with zero Zeeman
energy, the microscopic multi-skyrmion ansatz is presented. For
the system near $\nu=1$ with delta-function interaction, the bound
saturating BPS condition is shown to be satisfied by the
multi-skyrmion ansatz.  According to the parameters of the
explicitly constructed multi-skyrmion solutions, the dimension of
the moduli space of $k$ skyrmions is shown $4k+2$, corresponding
to the $2k$ positions, $k$ individual planar spin orientation, $k$
sizes, and $2$ the global spin axis(direction).

In section 4, we reformulate the system using the $*$-product
language to analyze the system studied in section 3. In section 5,
we derive in two ways the low-energy effective field theory for
the system: The $O(3)$ non-linear $\sigma$
model\cite{moon,Sondhi}, which describes the ferromagnetic spin
fluctuations in two-dimensions.  Its topological excitation is
known to be skyrmion \cite{Rajaraman}. The first method uses the
traditional many-body methods based on the Landau gauge. The
second method uses the noncommutative field theory which employs
the $*$-product. Comparing with the usual many-body approach, we
point out the usefulness of the noncommutative field theoretic
approach for the LLL physics.    Finally, section 6 is the summary
of our results.
\section{Lowest Landau Level systems and Noncommutative {$*-$}product}
\setcounter{equation}{0}
\renewcommand{\theequation}{\arabic{section}.\arabic{equation}}


We start with brief review of a charged particle state in a strong
magnetic field in terms of creation and annihilation operators.
In the symmetric gauge, the magnetic field is given by the vector
potential  $A_x=-{By/ 2},\ A_y= {Bx/ 2}$. We will choose the unit
of magnetic length $\ell = (\hbar c/e B)^{1/2}$ to be 1 in case
of no confusion. In terms of the complex coordinates $z=x+iy$ and
$\bar z=x-iy$, one can define two sets of oscillators:
\begin{eqnarray}
a=\sqrt{2}({\bar \partial} +{z} /4)&,&\
a^+=\sqrt{2}(-{\partial} +{\bar z}/4) \nonumber\\
b=\sqrt{2}({\partial} +{\bar z}/4)&,&\ b^+=\sqrt{2}(-{\bar
\partial} +{z}/4) \,,
\end{eqnarray}
where $\partial $ ($\bar \partial $)is a holomorphic
(anti-holomorphic) derivative satisfying $\partial \, z =1$ ($\bar
\partial \, \bar z =1 $) . The operators satisfy commutation
relations, $ [a, a^\dagger]=1 $, $ [b, b^\dagger]=1 $ and others
mutually commuting.

The Hamiltonian is described in terms of the first set of
oscillators, $a$ and $a^+$,
\begin{equation}
H= \hbar \omega_c(a^+a+\frac{1}{2})
\end{equation}
where $\omega_c = (m^* c / e B) $ is the cyclotron frequency. The
energy eigenvalues of the Landau levels (LL) are $E_n=\hbar
\omega_c (n+ 1/2 )$ with $n$ non-negative integer.

Each LL is degenerate and the degenerate state is distinguished in
terms of the second set of oscillators, $b$ and $b^+$, which
describe the guiding center coordinates.  Using the angular
momentum operator $L=2(b^+b-a^+a)$ which commutes with the
Hamiltonian $[L, H]=0$, one may assign quantum numbers to the
degenerate states: $|n,l\rangle$.  Since  $L$ satisfies the
algebra
\begin{eqnarray}
[L,b^+]=b^+ \,,\quad  [L,b]=-b
\end{eqnarray}
one raises (lowers) the angular momentum using $b^+$ ($b$) within
each LL.

The lowest Landau level is the set of states with $n=0$, which is
killed by $a$. The wave functions are of the form of $\psi(z)=
f(z) e^{- z \bar{z}/4}$ with $f(z)$ a complex analytic function
that does not grow too fast.  This analyticity is the difference
from that of the excited states, where $f$ is the arbitrary
function of $z$ and $\bar z$,  $f(z, \bar {z})$, in general.

The angular momentum eigenstate in the LLL as $|l \rangle =
\frac{1}{\sqrt{l !}} {b^\dagger}^l |0\rangle$ is given by
\begin{eqnarray}
 \langle z |l \rangle \equiv \Phi_l(z)
 =\frac{z^l}{\sqrt{2\pi
2^l l ! }}e^{-z\bar{z}/4}.\label{ang-mom-wavefunction}
\end{eqnarray}
This state forms thin shells of radius $\sqrt{2(l+1)}$ occupying
an area $2\pi$  since it is normalized as $ \langle z | z \rangle
= 1/2\pi$. The orbital degeneracy of a given Landau level is thus
$N_{\phi}=A/{2\pi}$, where $A$ is the total area.


Coherent state $|\zeta\rangle$, on the other hand, is the most
localized state in the LLL, which is defined as the eigenstate of
the annihilation operator $b$ as $\/ b|\zeta\rangle = {\bar
\zeta}|\zeta\rangle\/,$ or $\langle\zeta|b^\dagger = \langle\zeta|
{\zeta}$ .  The explicit construction is given as $|\zeta\rangle
= N_\zeta e^{\, {\bar \zeta} b^\dagger } |0\rangle$ where
$|0\rangle \equiv |0,0\rangle $ is the vacuum state. If we choose
the normalization constant $N_\zeta=
e^{-\frac{1}{2}\bar{\zeta}\zeta}/ \sqrt{2 \pi}$ so that $\langle
\zeta | \zeta \rangle =  1 / 2\pi$, then the coherent state has
the angular momentum component,
\begin{equation}\label{coherent}
  \langle \zeta | l \rangle =
\frac{1}{\sqrt{2 \pi l !}} \zeta^l
e^{-\frac{1}{2}\bar{\zeta}\zeta}\,.
\end{equation}
This allows one to identify the coherent state eigenvalue $\zeta$
with the coordinate $z$ for the LLL,
\begin{equation}\label{coord}
\zeta = {z  \over \sqrt{2}} \, ,\quad  {\bar \zeta} = {\bar z
\over \sqrt{2} }\,.
\end{equation}
Therefore, the LLL can be represented using either the angular
momentum basis or the coherent state basis.


The LLL is the projection of the Hilbert space into a subspace.
Any operator $\cal{O}$ acting on the larger Hilbert space of LL
can be similarly projected out so that the projected operator
denoted as $\hat{\cal{O}}$ acts only on the LLL states. This
projection is to put $a$-oscillators zero after normal ordering.
For example, a plane wave operator $e^{i{\bf p}\cdot{\bf r}}=
e^{2i(\bar{p}z+p\bar{z})} = e^{i\sqrt{2}(\bar{p}a +p a^\dagger)}
e^{i\sqrt{2}({\bar p}b^\dagger +{p} b)}$ becomes $e^{-{p}
\bar{p}}e^{i\sqrt{2}({\bar p} b^\dagger +{p} b)}$, where $p=(p_x +
i p_y)/2$ and  $\bar{p} =(p_x - i p_y)/2$.

One may choose any basis for the projected operators or their
matrix elements. Angular momentum basis is the simplest.
$|l\rangle\langle m |= :\frac{1}{\sqrt{l !}} {b^{\dagger}}^l
e^{-b^\dagger b} \frac{1}{\sqrt{m !}} b^{m}: $ ( $:~:$ is the
normal ordered product). On the other hand, the coherent state
basis provides a useful perspective for the non-commutativeness of
the projected operators. The matrix elements of the projected
identity operator with respect to the angular momentum states will
be $\delta_{mn}$. In the coherent basis, the corresponding
delta-function for the projected identity operator will be the
gaussian function :
\begin{equation}
\sum_l \langle \zeta |l \rangle \langle l |\zeta'\rangle ={1\over
2\pi}
\exp(-\frac{1}{4}|z|^2-\frac{1}{4}|z'|^2+\frac{1}{2}z\bar{z}')
\equiv \delta(z, z'). \label{proj-delta-fn}
\end{equation}
 Note that $b$ ($b^\dagger$) is realized as ${\bar \zeta}$
(${\zeta}$), and each forms one-dimensional subspace. However,
since $b$ and $b^\dagger$ do not commute, the two dimensional
space induces the noncommutativity and hence forms a
noncommutative plane \cite{read}.

It is well known that the multiplication of the non-commuting
operators can be realized in terms of the $*$-product of the
corresponding ordinary functions. The product of operators defined
on the noncommutative space of $b$ and $b^+$ is isomorphic to the
$*$-product of functions $f(\zeta)$.  One way of realizing this
isomorphism is to relate the normal ordered plane wave-like
operators $: e^{i({\bar p}b^\dagger +{p} b)}: $ to the plane wave
function $e^{-i(\bar p \zeta +  p \bar \zeta)}$. For the operator
defined by the Fourier transform \cite{read,pasquier},
\begin{equation}\label{*operator}
 \hat{O}_f(b,b^+ ) = \int { d^2p \over (2\pi)^2}
\tilde{f}(p, {\bar p } ) : e^{-i( {\bar p} b^+ + p b) } :
\end{equation}
the corresponding function $f(z,\bar z)$ is then given by
\begin{equation}
 f (\zeta,{\bar \zeta} )= \langle \zeta| \hat{O}_f(b,b^+ )| \zeta \rangle
 =\int { d^2p \over (2\pi)^2}
\tilde{f}(p, {\bar p } ) e^{-i( \bar p \zeta +  p \bar \zeta) } .
\end{equation}
The isomorphism of the $*$-product
\begin{equation}
\hat{O}_{f}\cdot \hat{O}_{g} = \hat{O}_{f* g}
\end{equation}
is easily verified if the $*$-product of functions is defined as
\begin{equation}
\left( f * g \right)(\zeta, \bar{\zeta})=e^{
\partial_{{\bar \zeta}}\partial_{{\zeta^\prime}}} f(\zeta, \bar{\zeta})
g(\zeta^{\prime}, \bar{\zeta^\prime}) |_{\zeta =\zeta^{\prime}}.
\end{equation}
If the function $f(\zeta , { \bar \zeta} )$ is defined in terms of
the complex coordinates $z$ and $\bar z$, then using the relation
Eq.~(\ref{coord}) we have
\begin{equation}
( f * g ) (z, {\bar z})= e^{ 2 {\bar \partial}  \partial^\prime }
f(z, \bar z ) g( z^{\prime }, \bar{ z^\prime } )  |_{z =
z^{\prime}} \,. \label{*product}
\end{equation}
This $*$-product is called $Q$-product in \cite{pasquier}. We will
just call this $*$-product.

This $*$-product enables one to write the expectation value of the
product of operators simply in terms of the corresponding function
\begin{equation}
\langle \zeta| \hat{O}_{f}\cdot \hat{O}_{g} | \zeta \rangle =
\langle \zeta| \hat{O}_{f* g}| \zeta \rangle
 = f*g (\zeta , { \bar \zeta} ) .
\end{equation}
The noncommutativity of $*$-product of functions is due to the
projection of the general states into the LLL.

We remark by passing that one may define different relations
between the operator and the function from Eq.~(\ref{*operator}),
which results in a different form of $*$-product. For example, by
using the Weyl ordering rather than the normal ordering, one may
have the so-called Moyal product where the front exponential
factor in Eqs.(\ref{*product}) will be given in terms of an
antisymmetric form.  In this article, $Q$-product is used since it
gives the natural functional relation for the spin system in
quantum Hall system.

\section{Skyrmion solutions: The BPS soliton}
\setcounter{equation}{0}
\renewcommand{\theequation}{\arabic{section}.\arabic{equation}}

We consider the quantum Hall system near $\nu=1$ with zero Zeeman
energy. In order to describe the many electron system in the LLL
with spin degeneracy, we introduce $\Psi$  operator which projects
to the LLL,
\begin{eqnarray}
\Psi =\sum_{l \sigma} |l,\sigma \rangle c_{l\sigma} \,.
\end{eqnarray}
The Hilbert space for this operator is the tensor product of
${\cal{H}}^{(1)}$  and  ${\cal{H}}_n$, {\it i.e.\/} the LLL state
and the occupation number states, respectively. The second
quantized field operator $\Psi_{\sigma}(z)$ projected down to the
LLL with spin $\sigma$ is written as
\begin{eqnarray}
\Psi_{\sigma}(z)=\sum_{l,\sigma'} \langle z, \sigma|l, \sigma'
\rangle c_{l\sigma'} = \sum_{l} \langle z|l \rangle c_{l\sigma}
\,.
\end{eqnarray}

The second quantized operator $c_{l\sigma}\ (c^\dagger_{l\sigma})$
which annihilates (creates) electron in the $l^{\rm th}$ orbital
with spin $\sigma$ satisfies the ordinary anticommutation
relations
\begin{eqnarray}
&&\{c_{l\sigma},c_{m\sigma'}\}=\{c^\dagger_{l\sigma},c^\dagger_{m\sigma'}\}
=0
\nonumber \\
&& \{c_{l\sigma},c^\dagger_{m\sigma'}\}=\delta_{lm}
\delta_{\sigma\sigma'} .
\end{eqnarray}
On the other hand, the field operator satisfies
\begin{eqnarray}
&&\{\Psi_{\sigma}(z),\Psi_{\sigma'}(z')\}=\{\Psi^\dagger_{\sigma}(z),\Psi^\dagger
_{\sigma'}(z')\} =0
\nonumber \\
&& \{\Psi_{\sigma}(z),\Psi^\dagger_{\sigma'}(z')\}=
\delta_{\sigma\sigma'}\delta(z,z')\,,
\end{eqnarray}
where the delta-function in $z$ has the gaussian form in
Eq.(\ref{proj-delta-fn}), not the usual two dimensional
delta-function, which reflects  the projection to the LLL. To
emphasize the projection to LLL we use only $z$ and no $\bar z$ in
$\Psi_\sigma$.

The field operator $\Psi_{\sigma}(z)$ ($\Psi^\dagger_{\sigma}(z)$)
annihilates (creates) an electron with a spin $\sigma$ at position
$z$ in the LLL. However, this interpretation is to be reserved
with the projected delta-function.  Only the operator $\Psi(z,\bar
z)$ ($\Psi^\dagger_{\sigma}(z,\bar z)$) with the full fermionic
annihilation (creation) operators $c_{(n,l)\sigma}\
(c^\dagger_{(n,l)\sigma})$ to all states $|(n,l), \sigma\rangle$
will satisfy the anticommutation rule with the true
two-dimensional delta-function and hence will annihilate (create)
the electron at the position $(z,\bar z)$. This projection results
in the noncommutative space and also becomes the source of
non-trivial electron excitation as we will see below.

In terms of the field operators $\Psi_{\sigma}(z)$ and
$\Psi^\dagger_{\sigma}(z)$, the Hamiltonian of the system is given
by
\begin{eqnarray}
\hat V= {1\over 2} \sum_{\sigma,\sigma^\prime} \int d^2 z \int d^2
z^\prime V(|z-z^\prime|)
\Psi^\dagger_{\sigma}(z)\Psi^\dagger_{\sigma^\prime}(z^\prime)
\Psi_{\sigma^\prime}(z^\prime) \Psi_{\sigma}(z)\,.
\end{eqnarray}
The ground state wavefunction for the completely filled $\nu=1$
LLL system is given by $|\Psi_0\rangle = \prod_{l}
c^\dagger_{l\uparrow}|0\rangle$, where all the spins are aligned
parallel to each other. In the coordinate basis, it can be
written by
\begin{eqnarray}
\langle z_1,..,z_N|\Psi_0\rangle=
\Psi_V(z_1,..,z_N)|\uparrow\uparrow
\uparrow\uparrow\cdots\uparrow\rangle . \label {groundwave}
\end{eqnarray}
where $\Psi_V$ is the Vandermonde determinant given by
\begin{eqnarray}
\Psi_V=\prod_{i<j} (z_i-z_j) \exp^{-\sum_k |z_k|^2/4}.
\end{eqnarray}

In free theory, any combination of $N$ states out of $2N$ states
of any spin configurations will all be degenerate. The Couloumb
interactions and the Zeeman splitting energy of the real system
will lift the above tremendous degeneracy. In the exact $\nu=1$
case, the ground state will still be the one in
Eq.(\ref{groundwave}). In the absence of Zeeman splitting, there
still exists extra spin degeneracy of $2 S_{\rm tot} +1$ with
$S_{\rm tot} = N / 2$. The system is called quantum Hall
ferromagnet. The lowest energy charge carriers of the system are
known to be skyrmions, which are spin-texture states and are
topological in nature. The following {\em variational} ansatz for
the skyrmion wavefunction was proposed by Moon {\em et
al.}\cite{moon}
\begin{equation}
\psi_\lambda = \prod_m \left(
    \begin{array}{c}
    z_m \\ \lambda
    \end{array}
\right) \Psi_V \label{skyrmionwavefn}
\end{equation}
When $|z|$ is smaller than $\lambda$, spins tend to align to the
$-{\hat z}$ direction. For $|z| \gg \lambda$, spins remain
aligned to the ${\hat z}$ direction, and so the parameter
$\lambda$ represents the size of the skyrmion.

In terms of the angular momentum basis $|m\rangle$, this skyrmion
wave function is in the form of
\begin{equation} |\Phi ; k \rangle =
\prod_m (\alpha_{m+k}c^\dagger_{{m+k}\uparrow} +\gamma_m
c^\dagger_{m\downarrow})|0\rangle ,
\end{equation}
where $k$ stands for the topological charge (winding number) of
the skyrmion. Here $k$ also represents the deficit of the number
of electrons from the completely filled $\nu=1$ state: $k = N -
N_e$.

For the delta-function interaction: $V(|z - z^\prime|)=V_0
\delta^2 (z - z^\prime)$, the Hamiltonian of the system is given
by
\begin{eqnarray}
\hat V= V_0 \int d^2 z
\Psi^\dagger_{\downarrow}(z)\Psi^\dagger_{\uparrow}(z)
\Psi_{\uparrow}(z) \Psi_{\downarrow}(z) \,.
\end{eqnarray}
This is a nonnegative operator with nonnegative energy for any
states. We will show that the skyrmion wavefunction
$\psi_\lambda$ saturates the inequality bounds of the energy. That
is, the skyrmion state satisfy the BPS condition of being
annihilated by the interaction operator ${\hat V}$:
\begin{equation} \label{BPSequation}
\Psi_{\uparrow}(z) \Psi_{\downarrow}(z)|\psi_\lambda\rangle=0.
\end{equation}
We impose the above BPS condition to the general state $|\Phi ; k
\rangle$ with $k=1$ and try to get the condition for $\alpha_m$
and $\gamma_m$. By explicitly writing down the field operator
$\Psi_{\sigma}(z)$ in terms of $c_{l\sigma}$, the annihilation
condition can be written by
\begin{eqnarray}
\sum_{l,m} \Phi_l (z) \Phi_m (z) c_{l\uparrow}
c_{m\downarrow}|\Phi ; 1\rangle ,
\end{eqnarray}
where $\Phi_m (z) = \langle z | m \rangle$ is the angular
momentum wavefunction in Eq.~(\ref{ang-mom-wavefunction}). The
coefficients $\alpha_m, \gamma_m$ should satisfy the following
relations
\begin{eqnarray}
{1 \over \sqrt{m+1}} {\alpha_{m+1} \over \gamma_m} = {\rm const}.
\label{BPS1}
\end{eqnarray}
We will calculate the spin profile of the state $|\Phi ; 1
\rangle$ satisfying Eq.(\ref{BPS1}). With the choice of the
constant to be $\sqrt{2} e^{i\theta_0} / \lambda$, the
expectation values of the spin operators in the state $|\Phi ; 1
\rangle$ are given by
\begin{eqnarray}
&& \langle S^\dagger (z) \rangle \equiv \langle \Phi ; 1 |
\Psi^\dagger_{\uparrow} (z) \Psi_{\downarrow} (z) | \Phi ; 1
\rangle = e^{-i\theta_0} {{\bar z} \over \lambda} \sum_m
|\gamma_m|^2 |\Phi_m (z)|^2
\nonumber \\
&& \langle S^{-} (z) \rangle \equiv \langle \Phi ; 1 |
\Psi^\dagger_{\downarrow} (z) \Psi_{\uparrow} (z) | \Phi ; 1
\rangle = e^{i\theta_0} {z \over  \lambda} \sum_m |\gamma_m|^2
|\Phi_m (z)|^2
\nonumber \\
&& \langle S_z (z) \rangle \equiv {1 \over 2}\langle \Phi ; 1 |
\Psi^\dagger_{\uparrow} (z) \Psi_{\uparrow} (z) -
\Psi^\dagger_{\downarrow } (z) \Psi_{\downarrow} (z) | \Phi ; 1
\rangle
\nonumber \\
&& \, \, \, \, \,= {1\over 2} \left( {|z|^2 \over \lambda^2}
-1\right) \sum_m |\gamma_m|^2 |\Phi_m (z)|^2 ,
\end{eqnarray}
where the coefficient $\gamma_m$ is determined by the
normalization condition: $\gamma_m = 1 / \sqrt{1 + 2
(m+1)/\lambda^2}$ \cite{Mac}. By normalizing the spin-density
fields over the unit area $2\pi \ell^2$ and multiplying by a
factor of $2$ to make the magnitude of the spin 1 : ${\bf m}=4\pi
\langle {\bf S} \rangle$, the magnetization field ${\bf m}$ is
given in the polar coordinates as follows
\begin{eqnarray}
&& m_x ({\bf r}) = {2r \over \lambda} \cos (\theta + \theta_0)
{\cal C}(r)
\nonumber \\
&& m_y ({\bf r}) = -{2r \over \lambda} \sin (\theta + \theta_0)
{\cal C}(r)
\nonumber \\
&& m_z ({\bf r}) = \left( {r^2 \over \lambda^2} -1\right) {\cal
C}(r) ,
\end{eqnarray}
where the function ${\cal C}(r)= 2\pi \sum_m |\gamma_m|^2 |\Phi_m
(z)|^2$. If we force to impose the normalization condition:
$|{\bf m}| = 1$, we can show that the function ${\cal C}(r)$ is
equal to $1 / [(r / \lambda)^2 +1 ]$ and ${\bf m}$ agrees {\rm
exactly} with the classical skyrmion solution for the $O(3)$
non-linear $\sigma$ model \cite{Rajaraman}. For example, with the
choice of $\theta_0=0$, the magnetization field ${\bf m}$ is
given by
\begin{eqnarray}
m_x ({\bf r}) = {2(x / \lambda)  \over (r / \lambda)^2 +1 },{}{}
m_y ({\bf r}) = -{2(y / \lambda) \over (r / \lambda)^2 +1 },{}{}
m_z ({\bf r}) = {(r / \lambda)^2 -1 \over (r / \lambda)^2 +1 } .
\end{eqnarray}
By calculating the function ${\cal C}(r)$ explicitly, one can
obtain the following result
\begin{equation}
{\cal C}(r)= \cases{1 / ( 1 + 2 /\lambda^2 ) & for $r \rightarrow
0$ \cr (r / \lambda)^{-2} & for $r > \lambda, \ell$} .
\end{equation}
For large skyrmions: $\lambda \gg \ell$, the function ${\cal
C}(r)$ indeed agrees with the classical solution. However, for
smaller ones, it deviates from the classical solution due to the
strong quantum fluctuations leading to $|{\bf m}| < 1$
\cite{kmoon3}. We notice that the skyrmion size and the relative
global spin orientations are determined by the choice of
$\lambda$ and $\theta_0$, and the skyrmion energy is
scale-invariant.

In the complex position space, the states $|\Phi ; 1 \rangle$
satisfying Eq.(\ref{BPS1}) are in the form of $\psi_\lambda$ in
Eq.~(\ref{skyrmionwavefn}). Hence, $\psi_\lambda$ is the {\em
exact} skyrmion solutions of the system, being the minimum of the
potential energy $\langle {\hat V} \rangle$ bounded from below by
$0$. The most general form of the skyrmion with unit charge will
be
\begin{equation}
\psi_1 = \prod_m \left(
    \begin{array}{c}
    z_m -\xi_1\\ a (z_m -\xi_2)
    \end{array}
\right) \Psi_V \, .
\end{equation}
We have three complex parameters for the moduli. The parameter $a$
is related to the background spin orientation of the skyrmion. The
center of mass value $\xi_c = (\xi_1+\xi_2)/2$ is related to the
location of the skyrmion, while the difference $\lambda =
(\xi_1-\xi_2)/2$ represents the size and the $U(1)$ planar
orientation. This can be easily seen from the fact that $\psi_1$
can be written in the similar form as in
Eq.~(\ref{skyrmionwavefn}) by making the spin rotation,
\begin{equation}
\psi_1 = \prod_m \left(
    \begin{array}{c}
    z_m - \xi_c\\ \lambda
    \end{array}
\right) \Psi_V'
\end{equation}
with respect to the globally rotated background spin.

Now we want to show that the multi-skyrmions are indeed BPS solitons.
The condition for the skyrmion to be a BPS soliton is that the
multi-soliton solutions are also annihilated by the interaction
${\hat V}$, Eq.~(\ref{BPSequation}). We study the two skyrmion
system. The following ansatz can correctly describe the two
skyrmion system, where one skyrmion is located at $z=a$ and the
other at $z=-a$
\begin{equation}
\psi_2 = \prod_m \left(
    \begin{array}{c}
    (z_m-a) (z_m+a) \\ \lambda^2
    \end{array}
\right) \Psi_V\,.
\end{equation}
When $a$ goes to zero, it reproduces a single $k=2$ skyrmion
wavefunction. For $a \gg \lambda$, it represents two
widely-separated skyrmions, whose size of each is about
$\lambda^2/(2a)$ and so it decreases with the distance $2a$. As
the distance $2a$ changes, the shape of the skyrmions are
distorted. We will show that in spite of the distortions of the
skyrmion shape, ${\hat V}$ still annihilates this two skyrmion
state. We begin with the class of general states $|\Phi ;
2\rangle$, which includes the two skyrmion states too
\begin{equation} |\Phi ; 2 \rangle =
\prod_{m=0} (\alpha_{m+2}c^\dagger_{{m+2}\uparrow} +\beta_m
c^\dagger_{m\uparrow} + \gamma_m c^\dagger_{m\downarrow} )
|0\rangle \label{twoskyrmion-ansatz}.
\end{equation}
Through the similar procedure to the single skyrmion case, we
obtain the following conditions for the coefficients $\alpha_m$,
$\beta_m$, and $\gamma_m$
\begin{equation}
{1 \over \sqrt{(m+1)(m+2)}} {\alpha_{m+2} \over \gamma_m} = 2 a_1
\end{equation}
and
\begin{equation}
{\beta_m \over \gamma_m} = a_2 ,
\end{equation}
where $a_1, a_2$ are arbitrary complex constants. In the complex
position space $z$, one can easily show that it is given by
\begin{equation}
\Phi_2 (z) = \prod_m \left(
    \begin{array}{c}
    a_1 z_m^2 + a_2 \\ 1
    \end{array}
\right) \Psi_V \,,
\end{equation}
which is the product of the Vandermonde determinant and the second
order polynomials. One can generalize the two skyrmion ansatz to
the multi-skyrmion ones, which can be written by
\begin{equation} |\Phi ; k \rangle =
\prod_{m} (\sum_{l=0}^{k_1} \alpha_{m+l}
c^\dagger_{m+l\uparrow} + \sum_{l=0}^{k_2}\gamma_{m+l}
c^\dagger_{m+l\downarrow} ) |0\rangle
\end{equation}
By imposing the BPS conditions to the above multi-skyrmion
ansatz\cite{pasquier2,moon}, the corresponding manybody
wavefunction in the $z$-space is written by
\begin{equation}
\Phi_k (z) = \prod_m \left(
    \begin{array}{c}
    f_{\uparrow} (z_m) \\ f_{\downarrow} (z_m)
    \end{array}
\right) \Psi_V \,,
\end{equation}
which again shows that it is the product of $\Psi_V$ and the
polynomial functions. The functions $f_{\uparrow} (z),
f_{\downarrow} (z)$ are polynomials of order $k_1$ and $k_2$,
respectively. The zeros of the polynomials represent the positions
of skyrmions, since the electrons are repelled by the zeros. Hence
they describe the manybody skyrmion wavefunctions with the
skyrmion number: $k=max [k_1, k_2]$.

Generically, the number of complex coefficients of $f_{\uparrow}
(z), f_{\downarrow} (z)$ are $2(k+1)$. One complex number is
related to the normalization of the wave function.  Hence we are
left with $4k+2$ real parameters for the multi-skyrmions with
skyrmion number $k$. Two real parameters from the ratio of the
coefficients of $z^k$ powers  in $f_{\uparrow} (z), f_{\downarrow}
(z)$ account for the global spin orientation in the asymptotic
region. For the well separated skyrmions, the remaining $4k$
parameters can be nicely interpreted as follows : We associate 4
real parameters for each skyrmion with two real parameters for the
position, one for the size and one for the $U(1)$ orientation.
This is consistent  with the fact that the multi-skyrmions are BPS
states with no interactioin energy.

Since any order of polynomials of the above form is annihilated
by ${\hat V}$, skyrmion is the BPS soliton.


\section{The $*$-product representation of the potential}
\setcounter{equation}{0}
\renewcommand{\theequation}{\arabic{section}.\arabic{equation}}


The potential energy of a state $|\Phi\rangle$ is written as
\begin{equation}
{\cal V} = \langle \Phi|\hat V |\Phi\rangle
 = \frac{1}{2}
 \int {\ d^2 z_1 } \int {\ d^2 z_2}~
 V(|z_1-z_2|) ~ \rho (z_1 | z_2)\,,
\label{energy}
\end{equation}
where $V (|z_1-z_2|)$  is the Coulomb potential of particles at
two points, $z_1$ and $z_2$. $\rho (z_1 | z_2)$ is the spin summed
density correlation
\begin{equation}
 \rho (z_1 | z_2)=
\sum_{\sigma_1,\sigma_2}\rho (z_1;\sigma_1 | z_2; \sigma_2)
\end{equation}
where the density correlation of a given spin configurations is
defined as
\begin{equation}
\rho (z_1;\sigma_1 | z_2; \sigma_2)
 =\langle \Psi^\dagger_{\sigma_1}(z_1) \Psi^\dagger_{\sigma_2}(z_2)
 \Psi_{\sigma_2}(z_2) \Psi_{\sigma_1}(z_1) \rangle \,.
 \label{density}
\end{equation}
Here $\langle \cdots \rangle$ represents the expectation value
with respect to a certain state $|\Phi\rangle$

As shown in the section 3, one may consider the simplest
delta-function interaction, which allows the existence of a
skyrmion solution.  The skyrmion solution is not very sensitive
to the shape of the potential and non-locality. So, we will
replace the Coulomb potential $V (|z_1-z_2|)$ with a
delta-function type potential and consider the skyrmion system in
the noncommutative formalism
\begin{equation}
 V (|z_1-z_2|) = V_0~\delta^2(z_1-z_2)\,.
\label{potential}
\end{equation}

The delta-function interaction simplifies the spin configuration,
since $\sigma_1$ and $\sigma_2$ contributes only when the two are
different.  Then the potential energy in (\ref{energy}) is written
as
\begin{equation}
{\cal V}  = \frac{V_0}{2}
 \int {{\ d^2 z }} ~ \rho (z_1 | z_2) |_{z_1 = z_2 =z}
\end{equation}
We will carry the index $\{ z_1, z_2\}$ of the density correlation
to avoid the possible confusion in the future.

Now the essence of the projection of the system to the LLL lies
in the density correlation operator (\ref{density}) where the
fermion operator is decomposed in terms of the LLL states only.
Therefore, to contrast the projected system with the full two
dimensional system we use the $*$-product to the correlation as
\begin{equation}
\rho(z_1;\sigma_1 | z_2; \sigma_2)
 =\langle (\Psi^\dagger_{\sigma_1}(z_1) \Psi^\dagger_{\sigma_2}(z_2))*
 (\Psi_{\sigma_2}(z_2) \Psi_{\sigma_1}(z_1)) \rangle \,.
 \label{*density}
\end{equation}

The $*$-product provides a convenient way of book-keeping of the
evaluation of operators, which reminds one of using the lowest
Landau level. What makes this $*$-product useful lies in the
possibility of presentation of the $*$-product in terms of
functions corresponding to this formal fermion operators. Then,
this $*$-product will provide a very simple way of finding an
effective potential for small fluctuations such as magnons with
respect to a certain background state.

The second quantized fermion operator cannot be simply represented
in terms of functions but its expectation value. To do this, we
start with the Hartree-Fock approximation of the density
correlation using the Wick theorem
\begin{equation}
\rho(z_1;\sigma_1 | z_2; \sigma_2)
 = 4\pi^2\left[d(z_1; \sigma_1| z_2; \sigma_2) - x(z_1 ;\sigma_1|z_2; \sigma_2)\right]\, ,
\label{HF}
\end{equation}
where $d(z_1;\sigma_1 | z_2; \sigma_2)$ and $x(z_1;\sigma_1 | z_2;
\sigma_2)$ are direct and exchange terms of the given spin
configuration respectively and are defined as
\begin{eqnarray}
d(z_1; \sigma_1 | z_2;\sigma_2) &=&
\langle\Psi^\dagger_{\sigma_1}(z_1) \Psi_{\sigma_1}(z_1)\rangle
\langle \Psi^\dagger_{\sigma_2}(z_2)
\Psi_{\sigma_2}(z_2)\rangle /4\pi^2\nonumber\\
x(z_1 ;\sigma_1| z_2;\sigma_2) &=&  \langle
\Psi^\dagger_{\sigma_1}(z_1) \Psi_{\sigma_2}(z_2)\rangle \langle
\Psi^\dagger_{\sigma_2}(z_2) \Psi_{\sigma_1}(z_1) \rangle /4\pi^2
\, ,
\end{eqnarray}
where the factor of $1/4\pi^2$ is attached since $\langle z| z
\rangle$ is normalized as $1/2\pi$. The potential
Eq.~(\ref{potential}) becomes
\begin{equation}
{\cal V}  = 4\pi^2 V_0  \int {{\ d^2 z }}  \left[ d(z_1 ;\uparrow|
z_2; \downarrow) - x(z_1;\uparrow | z_2; \downarrow)
 \right]|_{z_1 = z_2 =z}
\end{equation}
due to the spin exchange symmetry of the potential.  One may
consider the up-down spin configuration only as far as the
potential is concerned.  However, since this formalism is to be
extended to other physical quantity such as effective potential,
it is desired to consider all the spin configurations together.

The H-F potential is evaluated in the projected space and
therefore, this can be written in terms of the $*$-product. To
find the functional structure of the projected space one defines
operators, $D$ and $X$ corresponding to the functions $d$ and $x$,
which live only on the projected space $|z_1\rangle \otimes | z_2
\rangle $
\begin{eqnarray}
d(z_1 | z_2)
&=& \sum_{\sigma_1, \sigma_2} d(z_1;\sigma_1 | z_2; \sigma_2)\nonumber\\
& =& \langle z_1 |\otimes \langle z_2| \, D \,
|z_1\rangle \otimes | z_2 \rangle \nonumber\\
x(z_1 | z_2)
&=& \sum_{\sigma_1, \sigma_2} x(z_1;\sigma_1 | z_2; \sigma_2)\nonumber\\
 & = & \langle z_1 |\otimes \langle z_2| \, X \,
|z_1\rangle \otimes | z_2 \rangle \,.
\end{eqnarray}
Let us evaluate this $D$ and $X$ operators explicitly for
$|\Phi\rangle$,
\begin{equation} |\Phi; k \rangle =
\prod_m (\alpha_{m+k} \, c^\dagger_{{m+k}\uparrow} +\gamma_m \,
c^\dagger_{m\downarrow})|0\rangle \label{p-skyrmion}
\end{equation}
corresponding to the simplest skyrmion configuration ansatz with
topological charge $k$. $D$ and $X$ operators are expressed as the
spin summed operators:
\begin{equation}
  D= \sum_{\sigma_1 \sigma_2} D_{\sigma_1 \sigma_2}\,,\qquad
X= \sum_{\sigma_1 \sigma_2} X_{\sigma_1 \sigma_2}\,.
\end{equation}
The operator with given spin can be evaluated. We give explicit
representation for up-down spin configuration here.
\begin{eqnarray}
D_{\uparrow \downarrow} &=& \sum_{l,m} |l \uparrow\rangle \otimes
|m \downarrow\rangle {\bar \alpha_l} {\bar  \gamma_m}\, \alpha_l
\gamma_m  \langle l\uparrow|\otimes \langle m \downarrow|
\nonumber\\
X_{\uparrow \downarrow} &=& \sum_{l,m} |l \uparrow\rangle \otimes
|m \downarrow\rangle {\bar \alpha_l} {\bar \gamma_m} \,
\alpha_{m+k} \gamma_{l- k} \langle m+k \uparrow|\otimes \langle l-
k   \downarrow|
\end{eqnarray}
$\bar \alpha$ is the complex conjugate value of $\alpha$.

This $D$ and $X$ operators allow one to incorporate the
$*$-product easily.  For example, the direct term $D$ is written
as
\begin{equation}
  D= {\bar P_0} \,  P_0
\end{equation}
where
\begin{equation}
P_0 = \sum_{\sigma_1 \sigma_2} P_0^{\sigma_1 \sigma_2}\,.
\end{equation}
For the up-down spin configuration we have
\begin{eqnarray}
 P_0^{\uparrow \downarrow}  &=& \sum_{lm} (\alpha_l|l \uparrow  \rangle \langle l \uparrow| )
 \otimes ( \gamma_m \downarrow |m \rangle \langle m \downarrow|) \nonumber\\
&=& \sum_{lm} (|l \uparrow \rangle  \otimes |m\downarrow\rangle )
\alpha_l \gamma_m (\langle l \uparrow| \otimes \langle m
\downarrow |)\,,
\end{eqnarray}
and $\bar P_0$ is the hermitian conjugate.  This direct term is
written in terms of the diagonal projection operator $|l \rangle
\langle l|$ only. The exchange term $X$, however, needs
off-diagonal operator,
\begin{equation}
X= \bar P_0 \, Q_k
\end{equation}
where
\begin{equation}
Q_k = \sum_{\sigma_1 \sigma_2} Q_k^{\sigma_1 \sigma_2}\,.
\end{equation}
and
\begin{equation}
Q_k^{\uparrow \downarrow} = \sum_{l,m} (|l \uparrow\rangle \otimes
|m \downarrow\rangle) \alpha_{m+k}  \gamma_{l- k}(\langle m+k
\uparrow|\otimes \langle l- k \downarrow|) .
\end{equation}

Now it is clear that the H-F potential is written in terms of the
$*$-product of the function corresponding $P$ and $Q$,
\begin{eqnarray}
d(z_1 | z_2)  & =& \bar p_0 * p_0 \,(z_1 | z_2) \equiv e^{2
\bar\partial_1  \partial_1'}e^{2 \bar\partial_2  \partial_2'}p_0
(z_1 | z_2)~ p_0 (z_1' | z_2') \nonumber\\
 x(z_1 | z_2) & = & \bar p_0
* q_k \, (z_1 | z_2)
\equiv  e^{2 \bar\partial_1  \partial_1'}e^{2 \bar\partial_2
\partial_2'} p_0 (z_1 | z_2) ~q_k (z_1' | z_2')  \,.
\end{eqnarray}
where $*$-product is generalized into 2 coordinates from
Eq.~(\ref{*product}).

Of course, $p_0 $ and $q_k $ are defined as
\begin{eqnarray}
p_0 (z_1 | z_2)  & =& \langle z_1 |\otimes \langle z_2| \, P_0 \,
|z_1\rangle \otimes | z_2 \rangle
 \nonumber\\
q_k (z_1 | z_2)  & = & \langle z_1 |\otimes \langle z_2| \, Q_k \,
|z_1\rangle \otimes | z_2 \rangle \,.
\end{eqnarray}
Each function contains the spin sum. For the spin up-down
configurations, we have the following explicit form,
\begin{eqnarray}
p_0 (z_1 \uparrow| z_2 \downarrow) &=& \sum_{l,m} \alpha_l
\gamma_m |\langle z_1 | l \rangle |^2 ~|\langle z_2 | m \rangle
|^2\nonumber\\
q_k (z_1 \uparrow| z_2 \downarrow) &=& \sum_{l,m} \alpha_{m+k}
\gamma_{l-k} \langle z_1 | l \rangle \langle z_2 | m \rangle
\langle m+k |z_1 \rangle \langle l-k |z_2 \rangle  .
\end{eqnarray}

Now the H-F potential energy, Eq.~(\ref{HF}) is written as
\begin{equation}
{\cal  V} = 4\pi^2 V_0 \int {d^2 z}  \,{\bar p_0 } * ( p_0 - q_k
)\,(z| z)
\end{equation}

The potential energy has to be semi-positive definite. So, the
minimum of the H-F potential is obtained when $(p_0 -
q_k)(z\uparrow|z\downarrow)=0$.  The solution of this condition is
given as
\begin{equation}
 \frac{\alpha_{l+k}}{\gamma_l} \sqrt{\frac{l!}{(l+k)!}}
 ={\rm const}\,.
 \label{skyrmion-sol}
\end{equation}
This is the same BPS condition for the one skyrmion given in
Eq.~(\ref{BPS1}) from the  BPS equation in Eq.~(\ref{BPSequation})
\begin{equation}
 \Psi_{\downarrow}(z)\Psi_{\uparrow}(z)|\Phi\rangle=0 .\label{bpscondition}
\end{equation}

Hence, the skyrmion saturates the energy bound and is the BPS
soliton. From the non-commutative view point, the skyrmion is the
non-commutative soliton. The solution has the topological charge
$k$ if the coefficient satisfies the unitarity condition,
\begin{equation}
\label{unitarity}
  |\alpha_{l+k}|^2 + |\gamma_l |^2 =1 .
\end{equation}


\section{Non-linear $\sigma$ model: Manybody vs Noncommutative Field approach}
\setcounter{equation}{0}
\renewcommand{\theequation}{\arabic{section}.\arabic{equation}}


We derive the low-energy effective field theory for the $\nu=1$
quantum Hall system using the $|X\rangle$-basis obtained from the
Landau gauge: $A_x=0$, $A_y=Bx$. The real space wavefunction in
this gauge is given by
\begin{equation}
\Phi_X ({\bf r}) = {1 \over \sqrt{ \pi^{1/2} L_y }} e^{i k_y y}
e^{-(x-X)^2 / 2},
\end{equation}
where $X=k_y $ is the guiding center coordinate and $L_y$ the
length of the ${\hat y}$-dimension. The $|X\rangle$-basis is most
convenient in dealing with the $\hat x$-directional degrees of
freedom, while the $|m\rangle$-basis is for the radial degrees of
freedom. The Hamiltonian in the $|X\rangle$-basis is given by
\begin{equation}
{\hat V} = {1 \over 2} {\sum_{\sigma, \sigma^\prime}}
\sum_{X_1,X_2,X_3,X_4} V_{X_1 X_2 X_3 X_4} c^\dagger_{X_1 \sigma}
c^\dagger_{X_2 \sigma^\prime} c_{X_4 \sigma^\prime} c_{X_3 \sigma}
\end{equation}
where
\begin{equation}
V_{X_1 X_2 X_3 X_4} = \int d^2 z_1 \int d^2 z_2 V(|z_1 - z_2 | )
\Phi_{X_1}^* (z_1) \Phi_{X_2}^* ( z_2) \Phi_{X_4} (z_2)
\Phi_{X_3} (z_1).
\end{equation}
We use the following variational wavefunction $|\Psi\rangle$ for
the system\cite{moon},
\begin{equation}
|\Psi\rangle = \prod_X \biggl( c^\dagger_{X \uparrow} cos {\theta
(X) \over 2} + c^\dagger_{X \downarrow} sin {\theta (X) \over 2}
e^{i \phi (X)} \biggr) |0 \rangle .
\end{equation}
Since the state $|\Psi\rangle$ couples only the same $X$-orbitals
in the spin up and down states, it does not include charge
fluctuations. The energy expectation value for the state
$|\Psi\rangle$ is given by
\begin{eqnarray}
{\cal V}&=&\langle \Psi| {\hat V} | \Psi \rangle = {1 \over 2}
{\sum_{\sigma, \sigma^\prime}} \sum_{X_1,X_2,X_3,X_4} V_{X_1 X_2
X_3 X_4} \nonumber\\
&& \qquad \biggl[ \langle c^\dagger_{X_1 \sigma} c_{X_3 \sigma}
\rangle \langle c^\dagger_{X_2 \sigma^\prime} c_{X_4
\sigma^\prime} \rangle -\langle c^\dagger_{X_1 \sigma} c_{X_4
\sigma^\prime} \rangle \langle c^\dagger_{X_2 \sigma^\prime}
c_{X_3 \sigma} \rangle \biggr] .
\end{eqnarray}
One can obtain the following energy functional ${\cal V}({\bf m})$
in terms of the local magnetization field ${\bf m} (X)$,
\begin{equation}
{\cal V} ({\bf m}) = {1 \over 2} \sum_{X_1,X_2} \biggl\{ V_D - {1
\over 2} V_E \biggl( 1 + {\bf m} (X_1)\cdot {\bf m} (X_2) \biggr)
\biggr\},
\end{equation}
where $V_D$ stands for the direct interaction energy $V_{X_1 X_2
X_3 X_4}$ and $V_E$ the exchange energy $V_{X_1 X_2 X_3 X_4}$. We
calculate $V_D$ and $V_E$ for the case of the delta-function
interaction: $V (|z_1 - z_2 |) = V_0 \delta^2 (z_1-z_2)$, one can
obtain
\begin{equation}
V_D = V_E = V_0 \int d^2 z \left| \Phi_{X_1} (z )\right|^2 \left|
\Phi_{X_2} (z)\right|^2 = {V_0 \over \sqrt{2 \pi}} {1 \over L_y}
e^{-(X_1-X_2)^2 / 2 } .
\end{equation}
Putting in $V_D$ and $V_E$, we obtain the following energy
functional for the delta-function interaction
\begin{equation}
E ({\bf m}) = {V_0 \over {4 \sqrt{2\pi}}} {1 \over L_y}
\sum_{X_1,X_2} e^{-{(X_1-X_2)^2 / 2}} \biggl( 1 - {\bf m}
(X_1)\cdot {\bf m} (X_2) \biggr).
\end{equation}
As is clearly shown here, although we started with the {\em local}
delta-function interaction, our final energy functional becomes
{\em non-local}, which is due to the non-trivial dynamics at the
LLL space.

Assuming that the magnetization ${\bf m} (X)$ varies smoothly
over the magnetic length $\ell$, one can make a gradient
expansion and obtain
\begin{equation}
{\cal V} ({\bf m}) \approx {V_0 \over {32\pi^2}} L_y \int d X
\biggl( {{d {\bf m}} \over {d X}} \biggr)^2.
\end{equation}
Extending the above result to include the two-dimensional spin
fluctuations, we finally obtain the following $O(3)$ non-linear
$\sigma$ model\cite{Sondhi,moon}
\begin{equation}
{\cal V} ({\bf m}) \approx {1 \over 2} \rho_s \int d {\bf r}
\left( \nabla {\bf m} \right)^2 ,
\end{equation}
where the spin stiffness $\rho_s$ is given by $V_0/(16\pi^2)$.
The topological excitation of the non-linear $\sigma$ model is
known to be skyrmion excitation\cite{Rajaraman}. In section 3, we
described the microscopic skyrmion ansatz using the angular
momentum basis $|m\rangle$ and showed that they are the {\em
exact} BPS solitons of the system in the case of the {\em local}
delta-function interaction.

The above effective nonlinear $\sigma$ model can be obtained from
the noncommutative field theory.  Suppose most of the spins are
down and small perturbation for the up spins around the trivial
vacuum: $\alpha_l <<1$, and $\gamma_m \cong 1$ for $k=0$. Then
\begin{eqnarray}
p_0\, (z_1\uparrow | z_2 \downarrow ) &\cong&  {1 \over  2 \pi}
\sum_l a_l |\langle z_1 | l \rangle |^2 \nonumber\\
q_0\, (z_1\uparrow | z_2 \downarrow ) &\cong& \sum_{l,m} a_m
 \langle z_1 |l \rangle  \langle z_2 |m \rangle \langle m | z_1 \rangle
 \langle l | z_2 \rangle\,.
\end{eqnarray}
$p_0$ is the function of $z_1$ only and
\begin{equation}
\partial_1 q_0(z_1 \uparrow| z_2 \downarrow) |_{z_1 = z_2 =z} =0
\end{equation}
since
\begin{eqnarray}
&&
\partial_1 q_0 (z_1\uparrow | z_2 \downarrow )|_{z_1 = z_2 }
= \sum_{l,m} a_m \partial(
 \langle z |l \rangle  \langle m | z )\rangle
 \langle z |m \rangle   \langle l | z \rangle
  \nonumber\\
&&\qquad =\sum_{l,m} a_m (\partial
 \langle z |l \rangle \langle m | z \rangle
  \langle z |m \rangle\langle l | z \rangle  )
  -\sum_{l,m} a_m
 \langle z |l \rangle \langle m | z \rangle
 \partial   ( \langle z |m \rangle \langle l | z \rangle)
  \nonumber\\
&&\qquad = \sum_{m} {a_m \over 2 \pi}
 \partial (\langle z |m \rangle \langle m | z \rangle    )
  -\sum_{l,m} a_m
 \langle z |l \rangle \langle m | z \rangle
 (- {\bar z \over 2} + {m \over 2})\langle z |m \rangle
 \langle l | z \rangle
  \nonumber\\
&&\qquad = \sum_{m} {a_m \over 2 \pi}
\partial ( \langle m | z \rangle    \langle z |m \rangle)
  -\sum_{l,m} a_m
 \langle z |l \rangle \langle l | z \rangle
 \partial (\langle z |m \rangle \langle m | z \rangle   )
  \nonumber\\
&&\qquad = 0\,,
\end{eqnarray}
with $\sum_l \langle z| l \rangle \langle l |z \rangle = \langle
z| z \rangle = 1 / (2 \pi)$.  Now, the H-F density correlation
becomes,
\begin{eqnarray}
\rho (z | z ) &= &4\pi^2 {\bar p_0} * (p_0 - q_0) (z_1 | z_2)
 |_{z_1 = z_2 =z} \nonumber \\
 &=& 4\pi^2 e^{2\bar\partial_{1}\partial_1'}
 e^{2\bar\partial_{2}\partial_2'} {\bar p_0} (z_1 | z_2)
 (p_0 - q_0)(z_1' | z_2') |_{z_1 = z_2 =z_1' = z_2'=z} \nonumber \\
 & \cong & 8\pi^2 (\bar\partial \bar p_0^{\,\uparrow\downarrow } )
  (\partial p_0^{\,\uparrow\downarrow } )
\end{eqnarray}
where we use $p_0 (z| z) = q_0 (z| z) $ and the derivative
expansion for the long-wave length scale. In addition, the
magnetization $m_+ (z) \equiv  2 \langle \Psi^+ _{\uparrow } (z)
\Psi_{\downarrow} (z) \rangle $ is expressed as
\begin{equation}\label{m}
m_+ (z) \equiv  2 \langle \Psi^+ _{\uparrow } (z)
\Psi_{\downarrow} (z) \rangle \cong 4\pi \sum_l  a_l |\langle z |
l \rangle |^2  \equiv 8\pi^2 p_0 \,(z | z )\,.
\end{equation}
Therefore, we have the effective potential for the magnon field,
\begin{equation}
{\cal V} \cong  {V_0  \over 8\pi^2} \int d^2 z
 (\bar \partial  m_+ ~ {\partial }  m_-  ) \,.
\end{equation}
If one considers all the spin configurations, one has the
nonlinear $\sigma$ model for the effective action,
\begin{equation}
{\cal V} \cong { \rho_s \over  2} \int d {\bf r} ~4 \left(\bar
\partial {\bf m} \cdot {\partial }{\bf m} \right)= { \rho_s \over 2}
\int d {\bf r} \left(\nabla {\bf m}\right)^2 \, ,
\end{equation}
with the spin stiffness $\rho_s = { V_0/ ( 16 \pi^2)} $ and the
magnetization vector ${\bf m}$ normalized to $1$: $|{\bf m}|=1$ .

\section{Summary and Discussions}
\setcounter{equation}{0}
\renewcommand{\theequation}{\arabic{section}.\arabic{equation}}


We considered the quantum Hall system near the filling
factor $\nu=1$ with the negligible Zeeman splitting energy.  This
system possesses skyrmion collective excitations when projected to
the LLL with the Coulomb repulsion among electrons. Since its
presence is insensitive to the exact form of the potential, we
replace the Coulomb interaction with the delta-function
interaction.

We obtain the multi-skyrmion solutions, which are shown to be the
exact BPS solitons.  The system of skyrmions with total topolocal
charge $k$ has $4k+2$ moduli parameters. Two parameters
correspond to the global background spin orientation. For the
well separated skyrmions, four parameters can be associated to
each skyrmion: two for the location, one for the size, and one
for the $U(1)$ orientation. This is consistent with the fact that
they are BPS states.

The quantum fluctuation of the many electrons at LLL can be
described with the nonlinear $\sigma$ model, which is derived
using the traditional many-body approach. On the other hand, since
the system is projected to the LLL, one can study the system using
the noncommutative field theory approach. After writing the
Hamiltonian in terms of the noncommutative field theories, we
explicitly solve the BPS conditions for the excitation spectrum.
We also derive the low energy  effective nonlinear $\sigma$ model
action from the Hamiltonian through the $*$-product
representation. We have seen that this noncommutative field
approach gives clear insight into the property of the projection
as well as simplicity of the derivation and therefore, becomes
complementary to the many body approach. The noncommutative field
theory provides another natural and powerful framework in dealing
with the physics of the LLL states. Other physical properties of
the LLL system based on this language are under investigation.

\vskip 2mm

\noindent{\large\bf Acknowledgment} BHL and CR would like to
thank YVRC(Yonsei Visiting Research Center) for the warm
hospitality where a part of this work has been carried out. We
acknowledge financial support from the Basic Research Program of
the Korea Science and Engineering Foundation Grant No.
1999-2-112-001-5. The work of BHL is supported in part by BK21
Project No. D-0055.

\end{document}